\documentclass{iau}
\usepackage{graphicx}
\usepackage{graphics}
\usepackage{color}           
\usepackage{url}             
\usepackage{epstopdf}
\usepackage{float}

\title 
{Multi-wavelength Polarimetry and Variability Study of M87 Jet}

\author[Avachat etal.]   
{Sayali S Avachat$^1$, Eric S Perlman$^1$, William B Sparks$^2$, Mihai Cara$^{1,2}$, Frazer N Owen $^3$}

\affiliation{$^1$Deaprtment of Physics and Space Sciences, Florida Institute of Technology, \\150 W University Blvd, Melbourne, FL 32901 \\[\affilskip]
$^2$Space Telescope Science Institute,\\ 3700 San Martin Drive, Baltimore, MD 21218,\\[\affilskip]
$^3$National Radio Astronomy Observatory, Array Operations Center, \\P.O. Box No, 1003 Lopezville Road, Socorro, NM 87801-0387}

\pubyear{2014}
\volume{313}  
\pagerange{119--126}
\jname{Extragalactic Jets from Every Angle}
\editors{A.C. Editor, B.D. Editor \& C.E. Editor, eds.}
\begin{document}

\maketitle

\begin{abstract}
We present a high resolution polarimetry and variability study of the M87 jet using \it{VLA} and \textit{HST} data taken during 2002 to 2008. Both data-sets have an angular resolution as high as 0.06$"$, which is 2-3 times better than previous observations.  New morphological details are revealed in both the optical and radio, which can help to reveal the energetic and magnetic field structure of the jet. By comparing the data with previously published \textit{HST} and \textit{VLA} observations, we show that the jet$'$s morphology in total and polarized light is changing significantly on timescales of $\sim$a decade.  We compare the evolution of the inner jet (particularly the nucleus and knot HST-1), when our observations overlap with the multi-wavelength monitoring campaigns conducted with \textit{HST} and \textit{Chandra}.  We use these data to comment on particle acceleration and main emission processes.
\end{abstract}
\firstsection 
\section{Introduction}
\indent The radio galaxy M87 hosts one of the best-known extragalactic jets. Because of its proximity (d=16 Mpc, translating to a projected scale of about 80 pc per arcsec) and high surface brightness from radio through X-rays, studies of its jet emission can be undertaken at the highest resolutions in more wavebands than any other object. Its close proximity allows us to see structural detail in the jet$'$s structure very close to the nucleus, as well as changes on timescales of months to years. This has allowed much detailed work to be done on the physical processes at work in the jet. Previous observations of M87$'$s jet recorded a massive flare in knot HST-1, located 0.86$"$ from the nucleus (\cite[Harris et al. 2003]{Harris_etal03}, \cite[Harris et al. 2006]{Harris_etal06}, \cite[Harris et al. 2009]{Harris_etal09}, \cite[Perlman et al. 2003]{Perlman_etal03}, \cite[Perlman et al. 2011]{Perlman_etal11}, \cite[Madrid (2009)]{Madrid09}). During the flare, HST-1 increased in brightness by a factor of $\sim$100 in optical/UV (\cite[Perlman et al. 2011]{Perlman_etal11}, \cite[Madrid (2009)]{Madrid09}) and by even more in X-rays (\cite[Harris et al. 2009]{Harris_etal09}). The same observations showed month-timescale variability in both the nucleus and HST-1, as well as significant changes in the optical-UV spectrum and polarized emissions on similar timescales.  The spectral and polarization changes observed in HST-1 were found to be strongly correlated with the variability, indicating that the flare likely took place within a shock.\\
\indent Previous \textit{VLA} and \textit{HST} polarization observations of the jet taken in 1994-1995 were presented in \cite[Perlman et al. 1999]{Perlman_etal99}. They found a wide variety of polarized structures in the jet of M87, including evidence of shocks and jet instabilities.  Strong differences were found between the optical and radio morphologies in polarized light. \cite[Perlman et al. 1999]{Perlman_etal99} modeled these in terms of a stratified jet, in which high-energy, optical emitting particles were located closer to the jet axis than lower-energy, radio emitting particles.  Later work using deep \textit{Chandra} imaging data revealed these changes to be strongly correlated with the jet$'$s X-ray emission, thus revealing a deep connection between the magnetic field structure and high-energy emission( \cite[Perlman \& Wilson (2005)]{PerlmanWilson05}).\\
\indent We present new imaging polarimetry data taken with the \textit{VLA} and \textit{HST} during 2002-2008.  As these images represent an epoch approximately 10 years later than the work of \cite[Perlman et al. 1999]{Perlman_etal99}, they allow us to reflect back on that work.  Since these observations have an angular resolution 2-3 times better (0.06$"$-0.08$"$ as compared to 0.2$"$), they reveal new structural and morphological details.  In addition, both data-sets cover the time domain of the flare of HST-1. We examine the evolution of the polarized structure of that region of the jet during the flare.  We can observe structural changes in the innermost region of the jet, with new components forming and moving downstream of the nucleus and HST-1. Other structural details are also visible downstream. Comparison of the polarization structure in different wavelengths can help us understand the physics behind shock acceleration and the implications in the case of the flaring of HST-1, and the observed variability of the nucleus.
\section{Results and Discussion}
\indent During 2002-2008, M87 was observed by the VLA every 5-6 months at 8, 15, 22 and 43 GHz in A and B configurations. \textit{HST} observations were obtained about every month during the same time-period (\cite[Madrid (2009)]{Madrid09}) with the polarimetry mostly restricted to the time interval between late 2003 and late 2006 (\cite[Perlman et al. 2011]{Perlman_etal11}). \textit{VLA} data were extracted from the \textit{NRAO} Data Archive (https://archive.nrao.edu), while the \textit{HST} data were extracted from the \textit{HST} archive (http://archive.stsci.edu).
\indent Both the optical and radio data were reduced using standard techniques.  These are detailed in \cite[Perlman et al. 2011]{Perlman_etal11} for the \textit{HST} data, and we will detail these for the \textit{VLA} data in a later paper.  As the optical data were taken at a wavelength of approximately 6000 \AA (using ACS/HRC and the F606W+POLVIS filters), they have a diffraction limited angular resolution of 0.06$"$. As the K band (22 GHz) data have the highest signal to noise of all the \textit{VLA} data, in this paper we use only those data for the comparison of the jet$’$s optical and radio polarized structure. \\
\indent Fig.1 shows the 2005 polarization images at 22 GHz and F606W as compared to radio and optical images from \cite[Perlman et al. 1999]{Perlman_etal99}. The magnetic field polarization vectors are overlaid onto the total flux contours in each. The differences such as the better resolved nucleus and HST-1 regions are clear in our maps as compared to the older maps in \cite[Perlman et al. 1999]{Perlman_etal99}. The new maps show significant differences as compared to the older images. We see components close to the nucleus and HST-1 are more resolved than previous images. In addition, knot HST-1 is much brighter in the 2005 images than in 1994-1995, and it is also resolved into multiple components.  Another interesting region in the inner jet, knot D, is observed to have more complex structure in the new high-resolution radio and optical images.\\
\indent Both images in Fig 2. (left hand panel) show quite similar polarization structure in the region of the nucleus. The magnetic field vectors are mostly aligned with the direction of the jet axis with uniform polarization and seem to rotate along the flux contours near the jet edge. HST-1 is much more highly polarized ($\sim$20\%) than the nucleus ($\sim$3-4\%) in both wavebands. The magnetic field vectors in HST-1 predominantly lie perpendicular to the jet axis in the optical, indicating a strong shock.  By comparison, in the radio they lie along the jet axis and, similar to the nucleus, seem to rotate along the flux contour near the surface.  The very different radio and optical polarization morphologies are further evidence of a stratified energetic structure with high-energy optical to X-ray emitting particles located closer to the jet axis, as suggested in \cite[Perlman et al. 1999]{Perlman_etal99}. This indicates that the shock associated with HST-1 is located within the jet interior.\\
\indent In the 22GHz image, there seem to be a few new components emerging out of the nucleus and HST-1 in our radio maps. The stacked radio map shows the components emerging downstream of HST-1 out up to 1.5$"$ from the nucleus. Such similar superluminal components forming and moving downstream of HST-1 were reported in \cite[Cheung et al. 2007]{Cheung_etal07} using their \textit{VLBA} observations in which the fastest moving component, HST-1a, had a speed of about 2.5c. It is argued there, that the formation of these superluminal components may be related to the observed variability of the nucleus and HST-1. It will be interesting to combine the \textit{VLBA} observations with our \textit{VLA} observations and analyze the jet based on variability of superluminal components. \\
\indent Fig.2 (right hand panel) shows a similar comparison of components in the outer jet, namely knot A and B. The optical and radio polarization observations show complex position angle structure in the interior of the knot A and B complex and show parallel vectors along the edges. This is consistent with the model suggesting the lower energy radio electrons are most possibly located near the surface of the jet. 
We do not see any evidence of correlation between the radio flux and polarization in this region. Both the knots are highly polarized indicating a highly ordered magnetic field. In optical we see a strong correlation between flux and polarization. The maxima (and minima) of the polarization coincide with the maxima (and minima) of the total flux. As a result of this variation, we can trace at least 4 wrappings of a helix formed by the optical magnetic field vectors in knot A and B maxima. In the high polarization region upstream of knot A, we see that the magnetic field vectors are aligned perpendicular to the direction of jet flow. This is the brightest part of the wide knot A, called A shock in the literature (e.g. \cite[Perlman et al. 2001]{Perlman_etal01}) and is assumed to be the region responsible for particle in situ acceleration. We see a similar magnetic field morphology in radio image.\\
\indent Fig.3 shows total flux, percent polarization and electric vector position angle (EVPA) plotted over the period of observations. The fractional polarization of the nucleus was lower but  stayed constant within a few sigma. The polarization of the nucleus did not show any evidence of correlation with the observed variability of nucleus in any band. For the HST-1, the value of polarization was much higher, (plot not shown here) but similar to the nucleus it stayed constant within a few sigma with no correlation to the flux variability. In the optical, on the other hand, (\cite[Perlman et al. 2011]{Perlman_etal11}), the HST-1 showed very different behavior. In optical, the polarization was observed to increase to more than 40\% during its flare and a sharp decrease in its value as the flux decreased.\\
\indent The EVPA of the nucleus and HST-1 (plot not shown here) stayed constant throughout the radio observations. Its value was about 70 degrees on an average, which is close to the nominal jet PA. This behavior of EVPA of HST-1 was similar in both the radio and the optical but was very different for the nucleus. In the optical, EVPA of the nucleus varied more than 120 degrees during the period of observations.\\
\indent The plot of total flux of the nucleus (plot not shown here) indicated constant nuclear flux throughout the period of observations. The optical data during the same time period showed month time scale variability in the nuclear flux along with two small flares in the nucleus, one in early 2005 and another in early 2007 (\cite[Perlman et al. 2011]{Perlman_etal11}). We did not see month time scale flux variability in the nucleus because the observations were taken every 4-6 months, but the small increase in flux in mid-2007 may have been caused due to flaring activity of nucleus. The total flux versus time plot for HST-1 showed its flaring activity. The brightness of the knot gradually increased from late 2002 and was at its peak brightness in mid-2005. After that there was a rapid decrease in its brightness and no other flare or variability was observed during the remaining time of observations until early 2008. The observed increase in radio flux was relatively smaller as compared to the optical where the increase was a factor of 100.
\section{Conclusion}
We presented our results of multi-wavelength polarimetry study of the M87 jet. Our data show very interesting results in terms of the variability of various components along the jet. There were significant structural changes in the components and the high resolution data allowed us to see these changes on the sub arc-second scales. Polarimetry comparison of the multi-wavelength data-sets further allow us to probe into particles located in different regions of the jet. High resolution data spanning over several years can give us a wealth of information about the changes in the magnetic field structure and its role in particle acceleration mechanisms. We further plan to add \textit{VLBA} and X-ray observations to this multi-wavelength variability study.

\begin{figure}[H]
\begin{center}
  \begin{tabular}{@{}cc@{}}
  \includegraphics[width=4cm,height=6.7cm, angle=270]{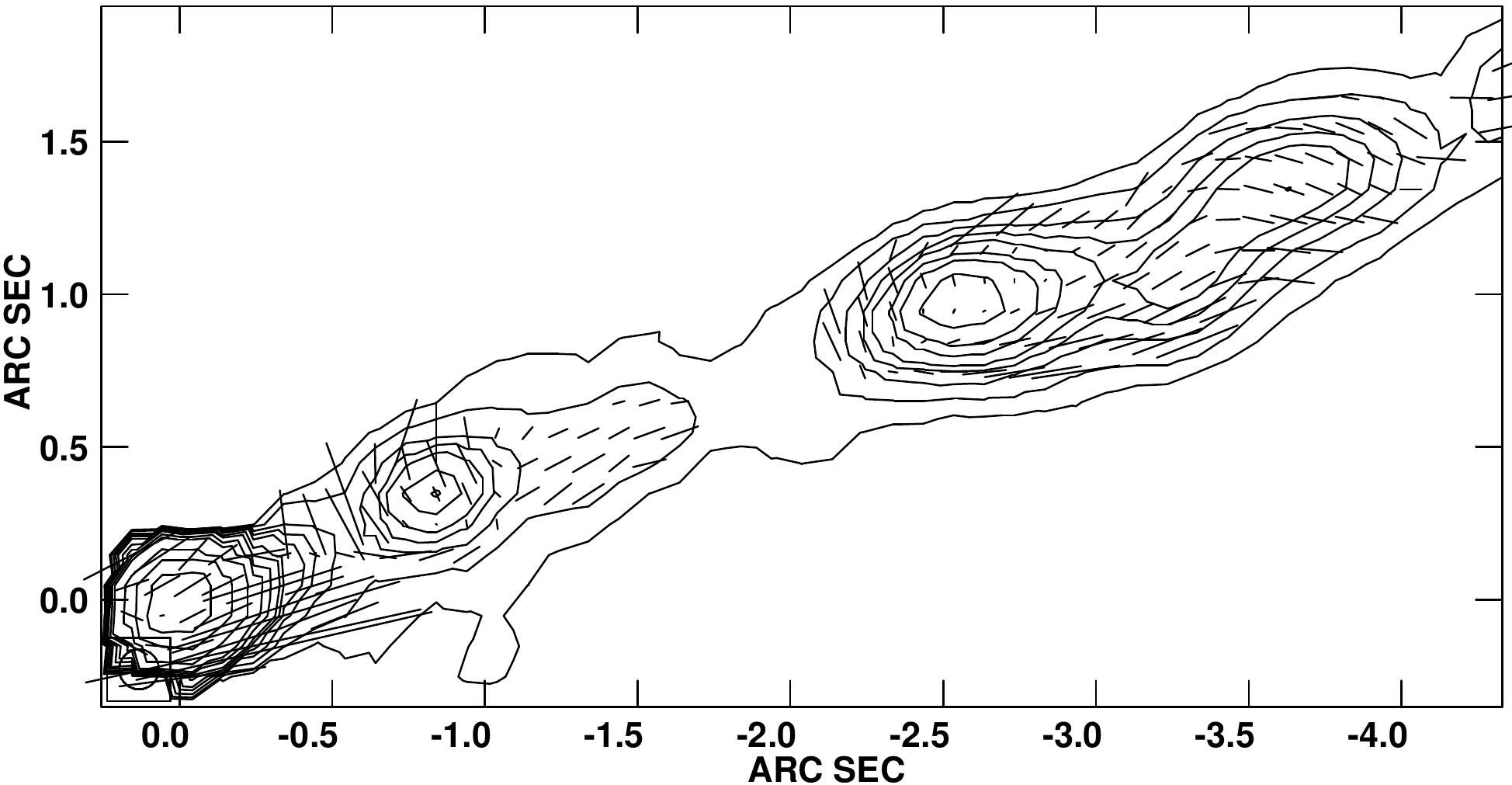}
  \includegraphics[width=4cm,height=6.7cm,angle=270]{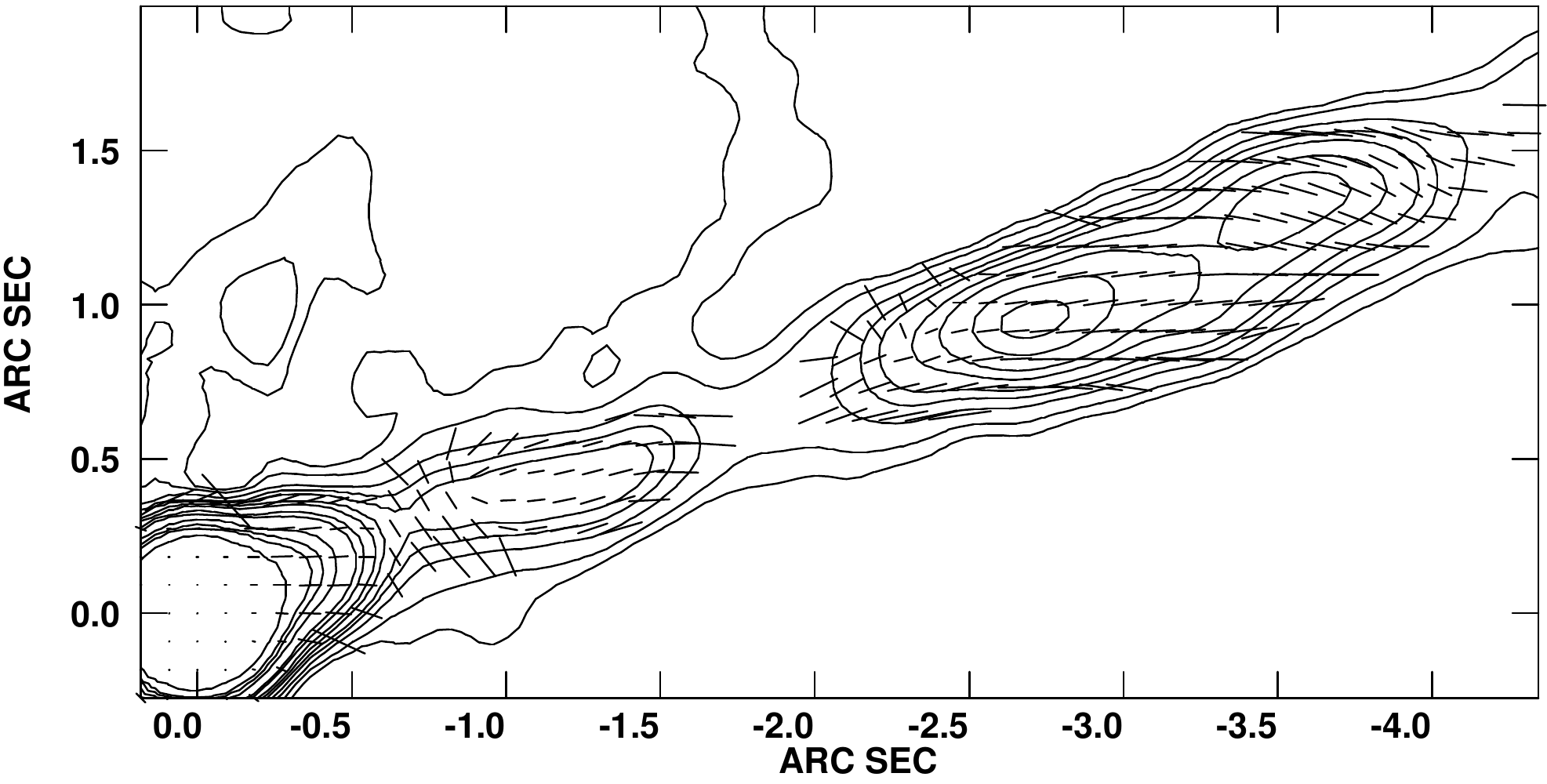}\\
  \includegraphics[width=6.7cm,height=4cm]{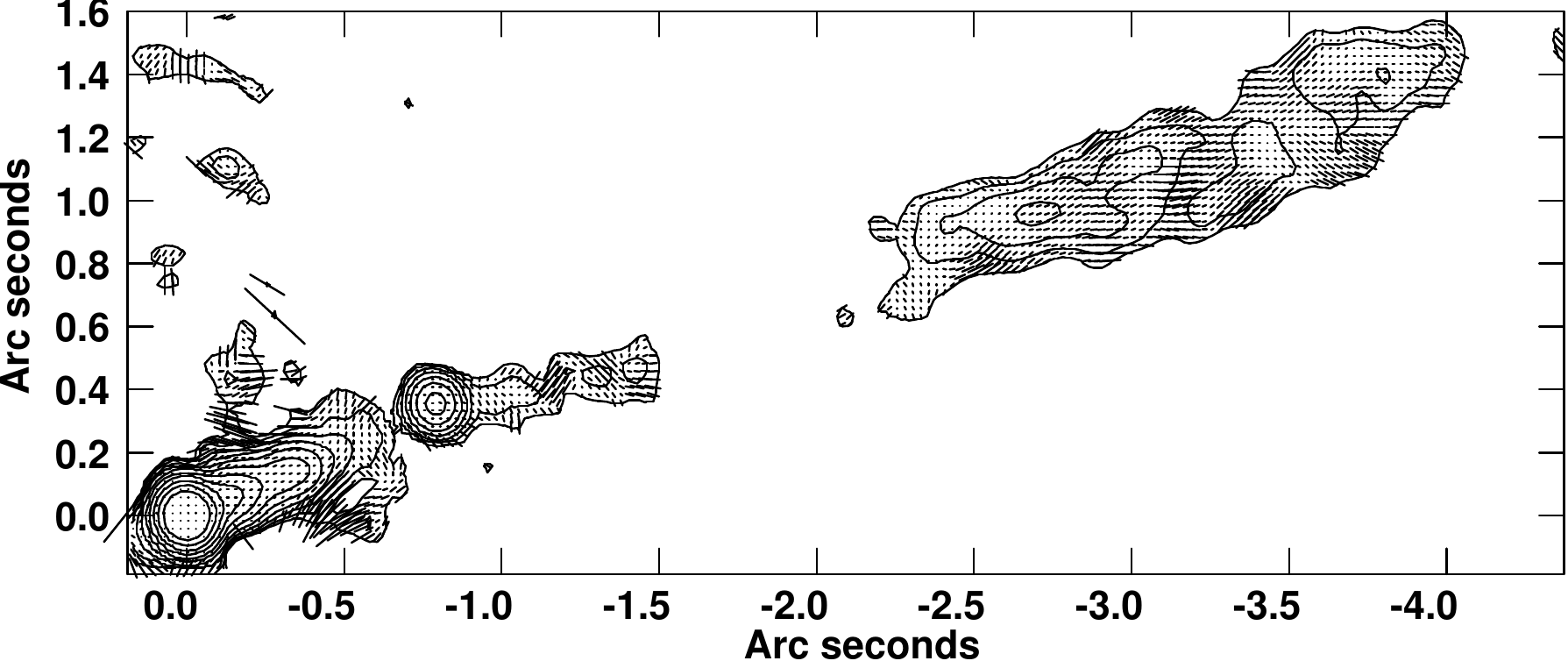}
  \includegraphics[width=6.7cm,height=3.9cm]{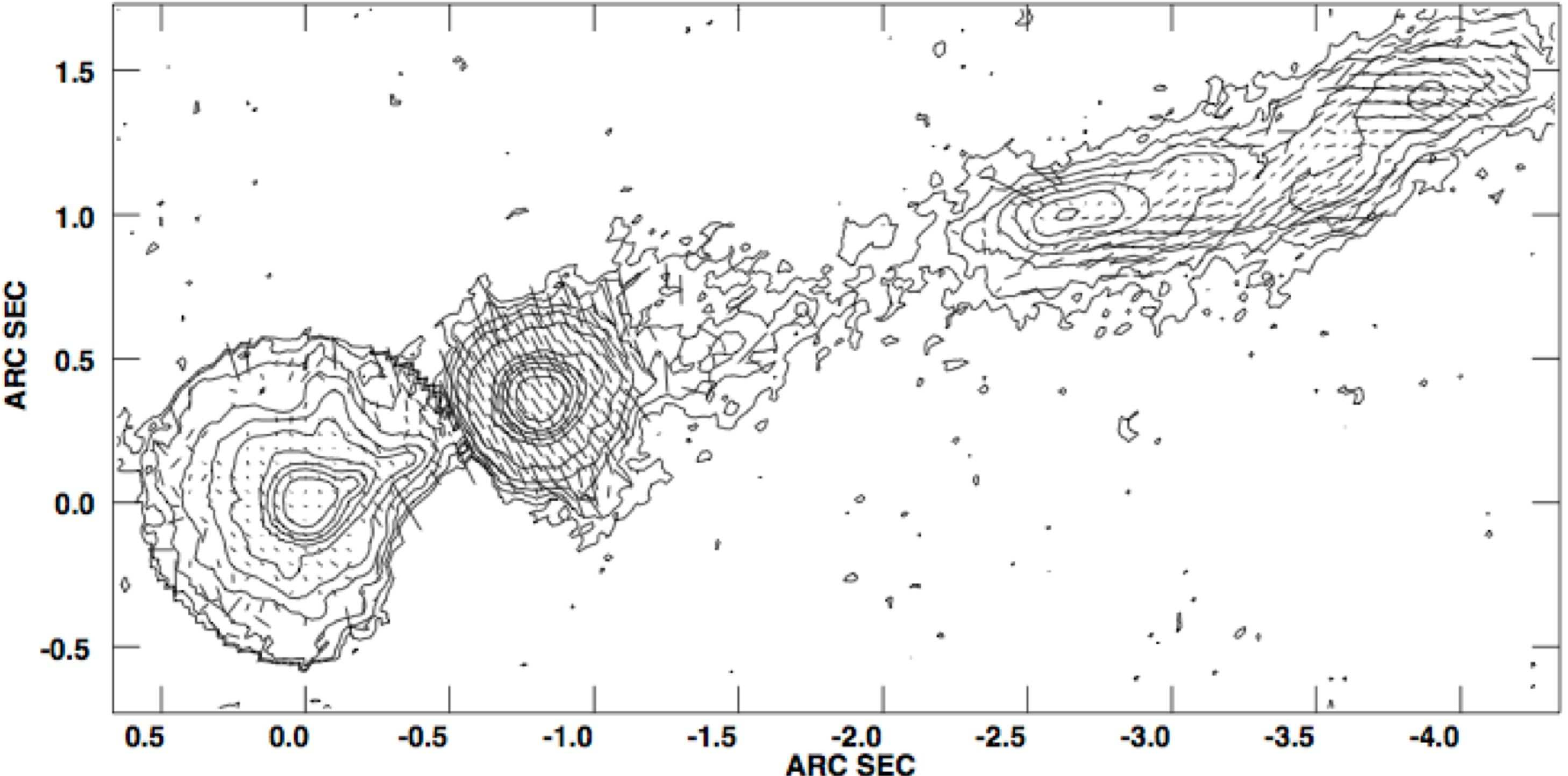}\\
\end{tabular}
\caption{Comparison of the radio and optical polarimetry: Top - The radio (15 GHz) and optical polarization (F555W) data from 1994 (\cite[Perlman et al. 1999]{Perlman_etal99}); Bottom - The radio (stacked 22 GHz) and optical polarization (stacked F606W) between 2002-2008}
  \end{center}
  \label{fig:1}
\end{figure}
\begin{figure}[H]
\begin{center}
 \includegraphics[width=6.7cm, height=4cm]{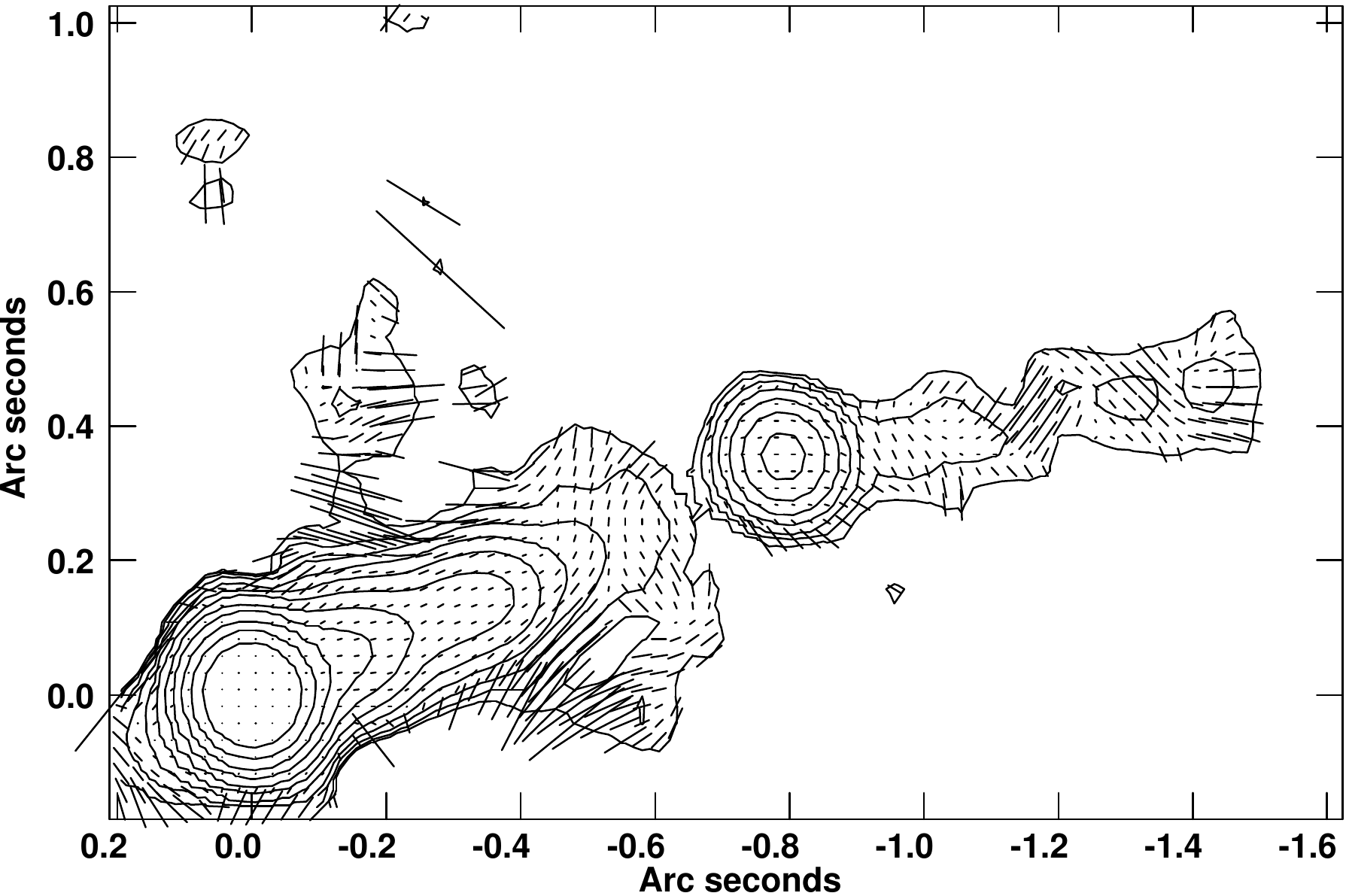}
 \includegraphics[width=6.7cm, height=4cm]{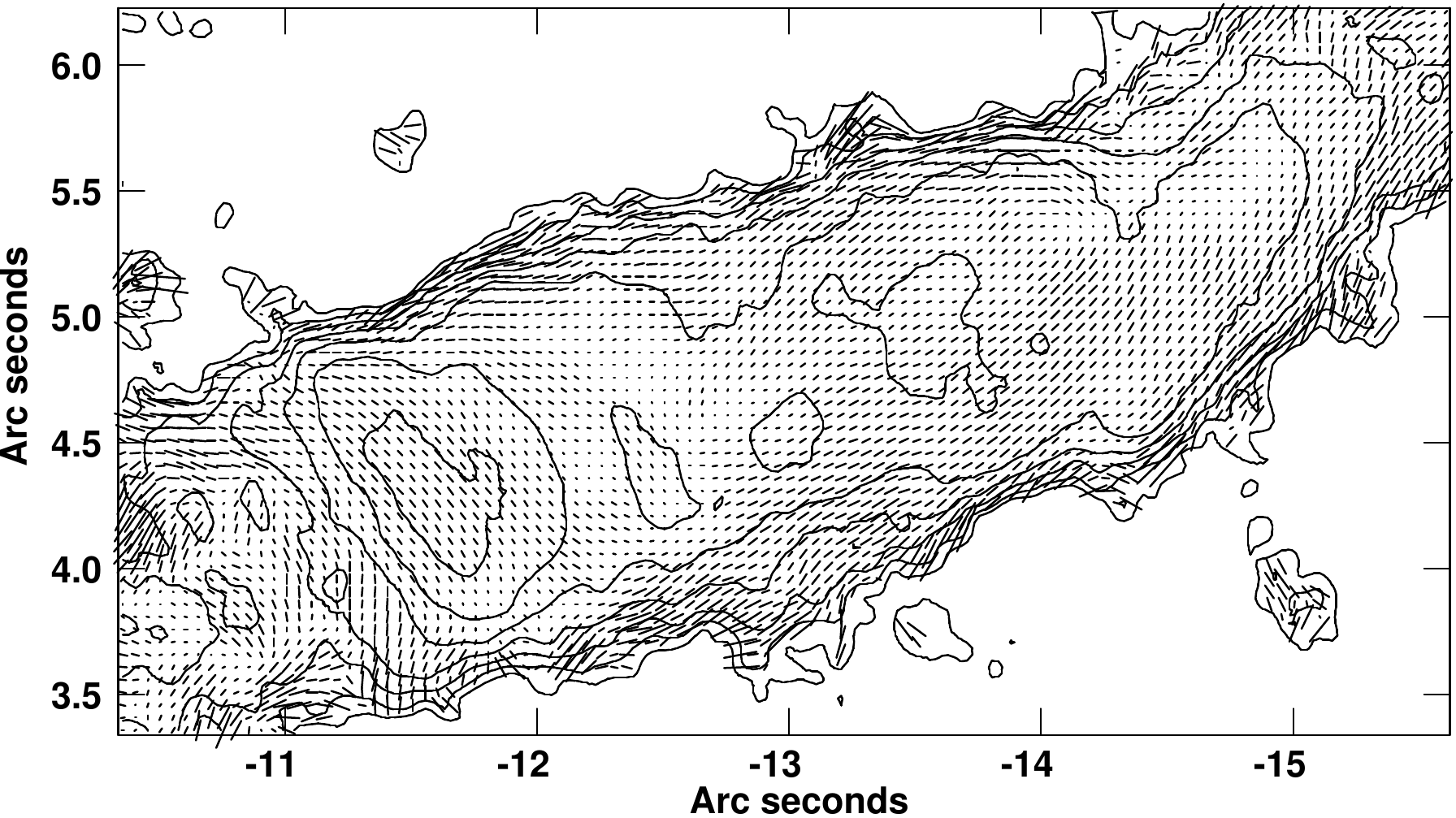}\\
 \includegraphics[width=6.7cm, height=4cm]{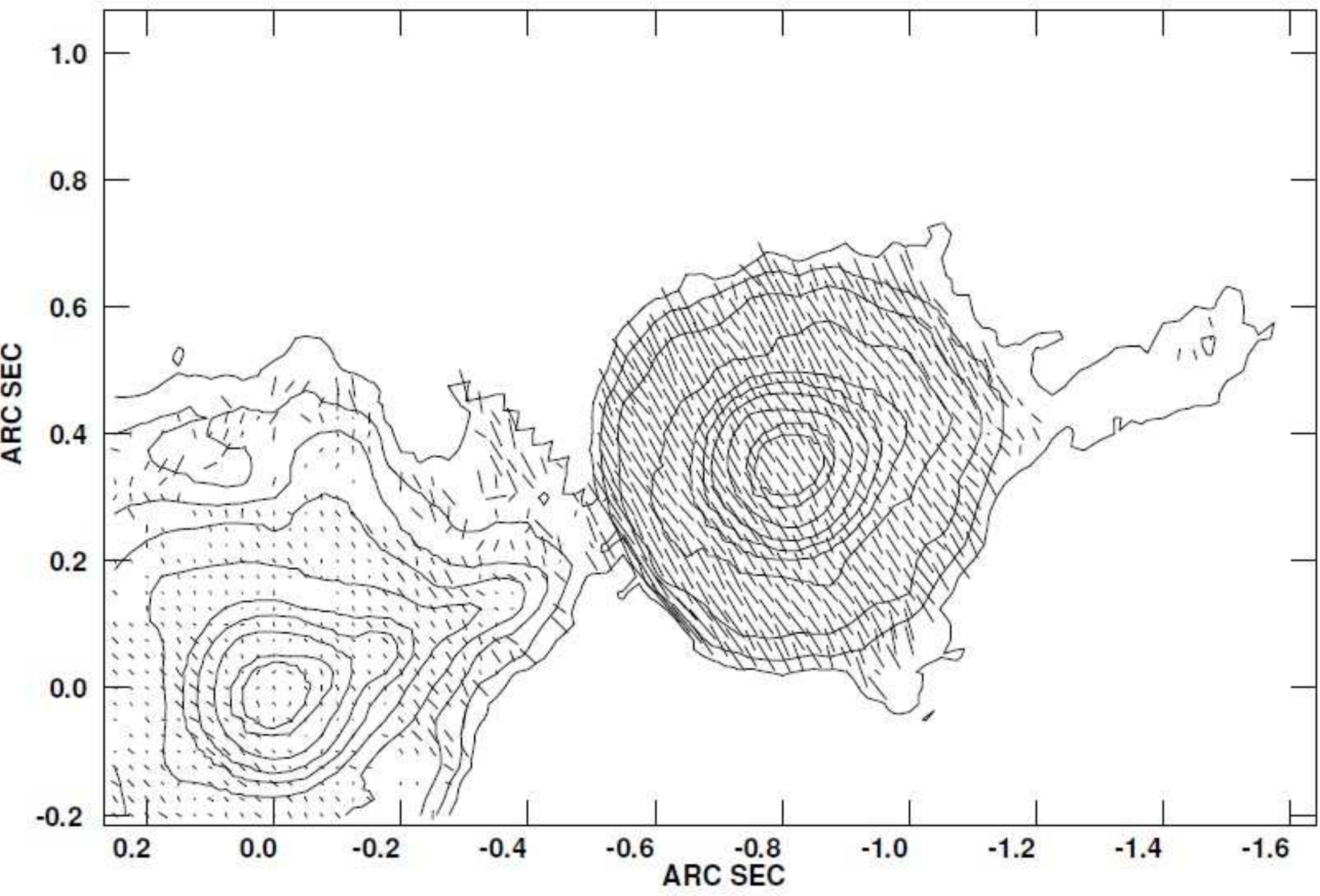}
 \includegraphics[width=6.7cm, height=4cm]{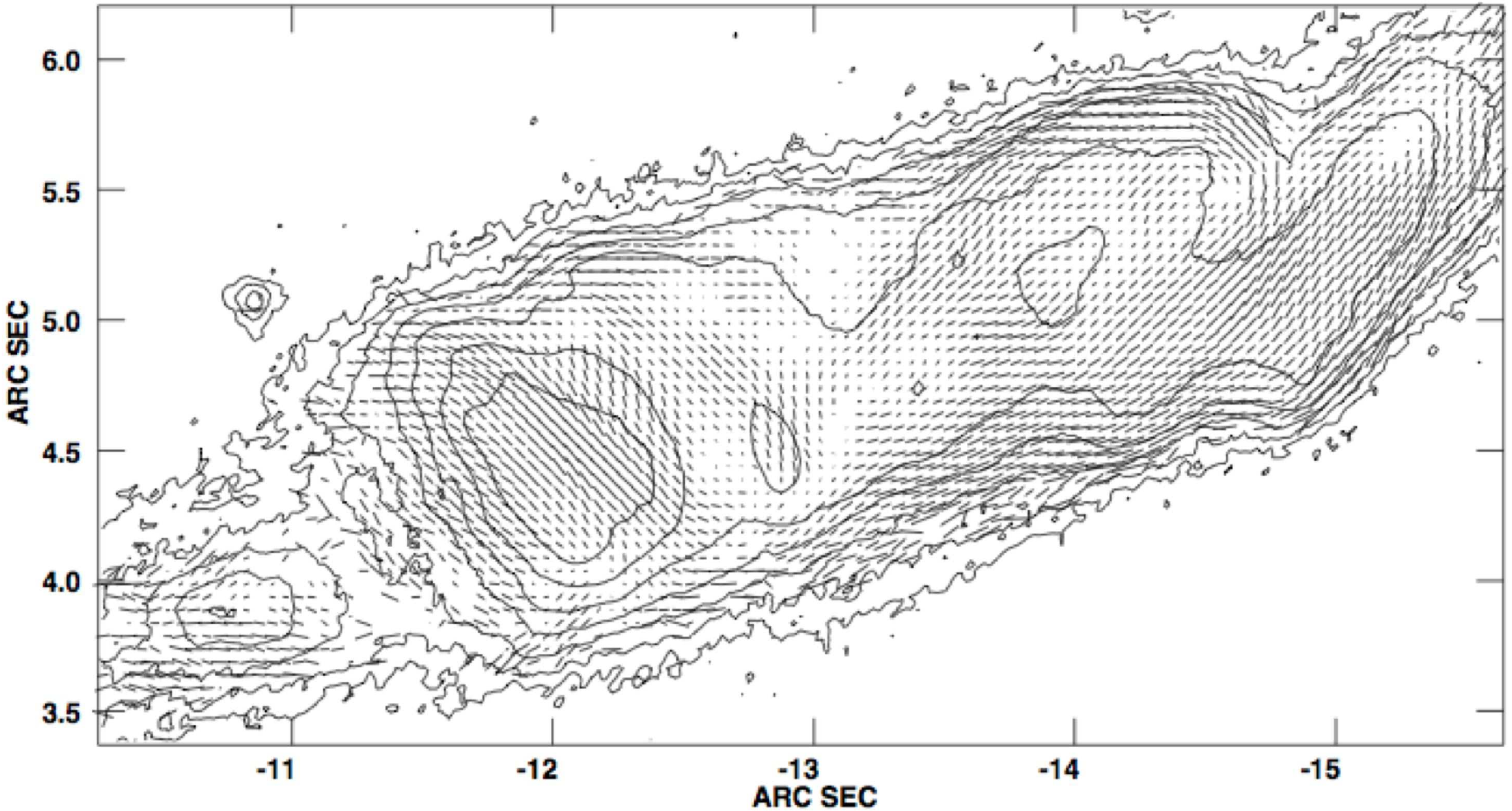}
 \caption{Right - The magnetic field polarization vectors overlaid by the total flux contours of the innermost jet showing the nucleus and HST-1. Top - The 22GHz stacked images of composite A and B array show, the nucleus and HST-1 (on left) and knots I, A and B (on right) Bottom - The Stacked optical images at wavelength F606W show the nucleus and HST-1 (on left) and knots I, A and B (on right)}
    \end{center}
   \label{fig:2}
\end{figure}
\begin{figure}[H]
  \begin{center}
  \begin{tabular}{@{}cc@{}}
   \includegraphics[width=6.3cm, height=4.5cm]{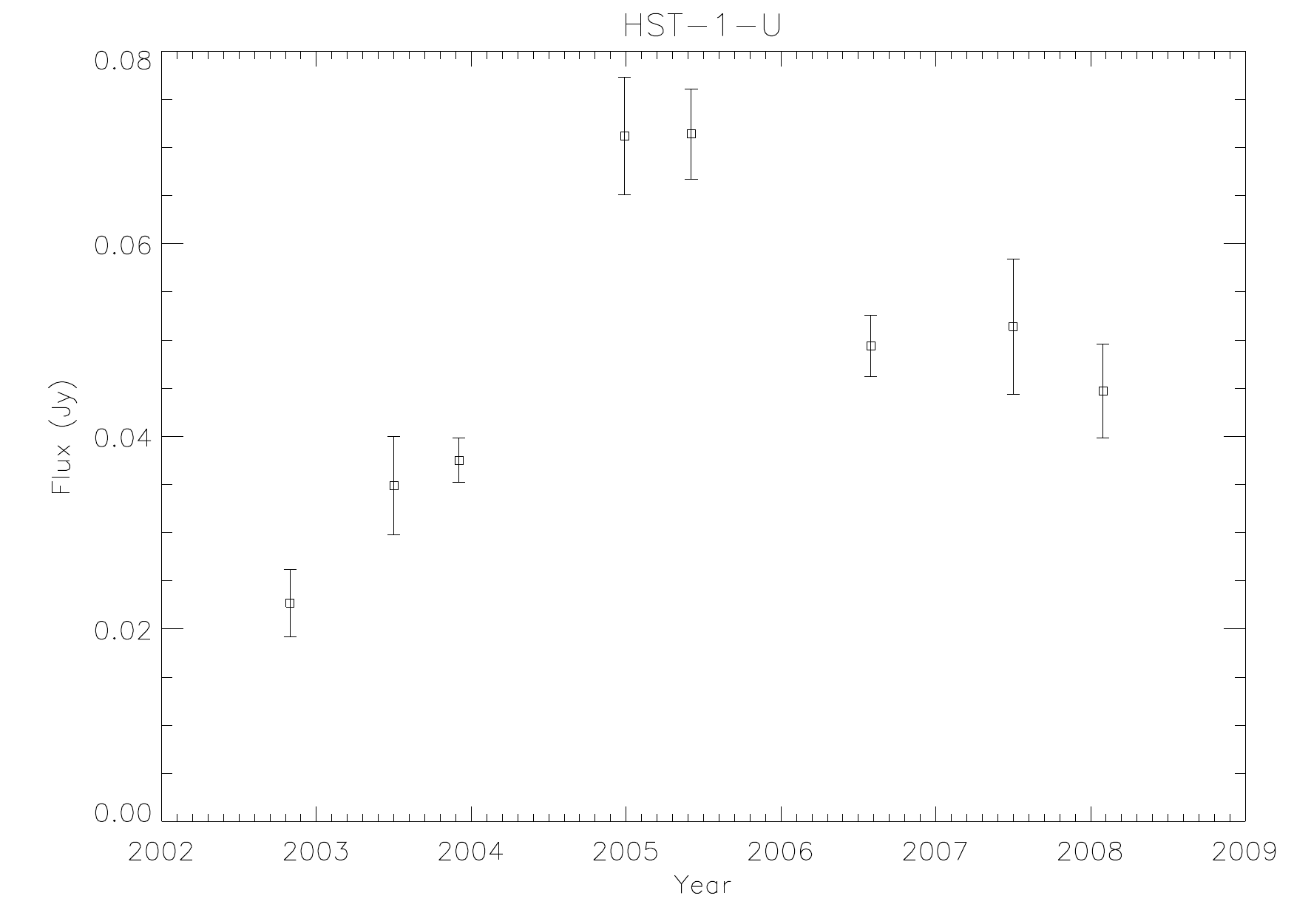}
    \includegraphics[width=6.9cm, height=4.5cm]{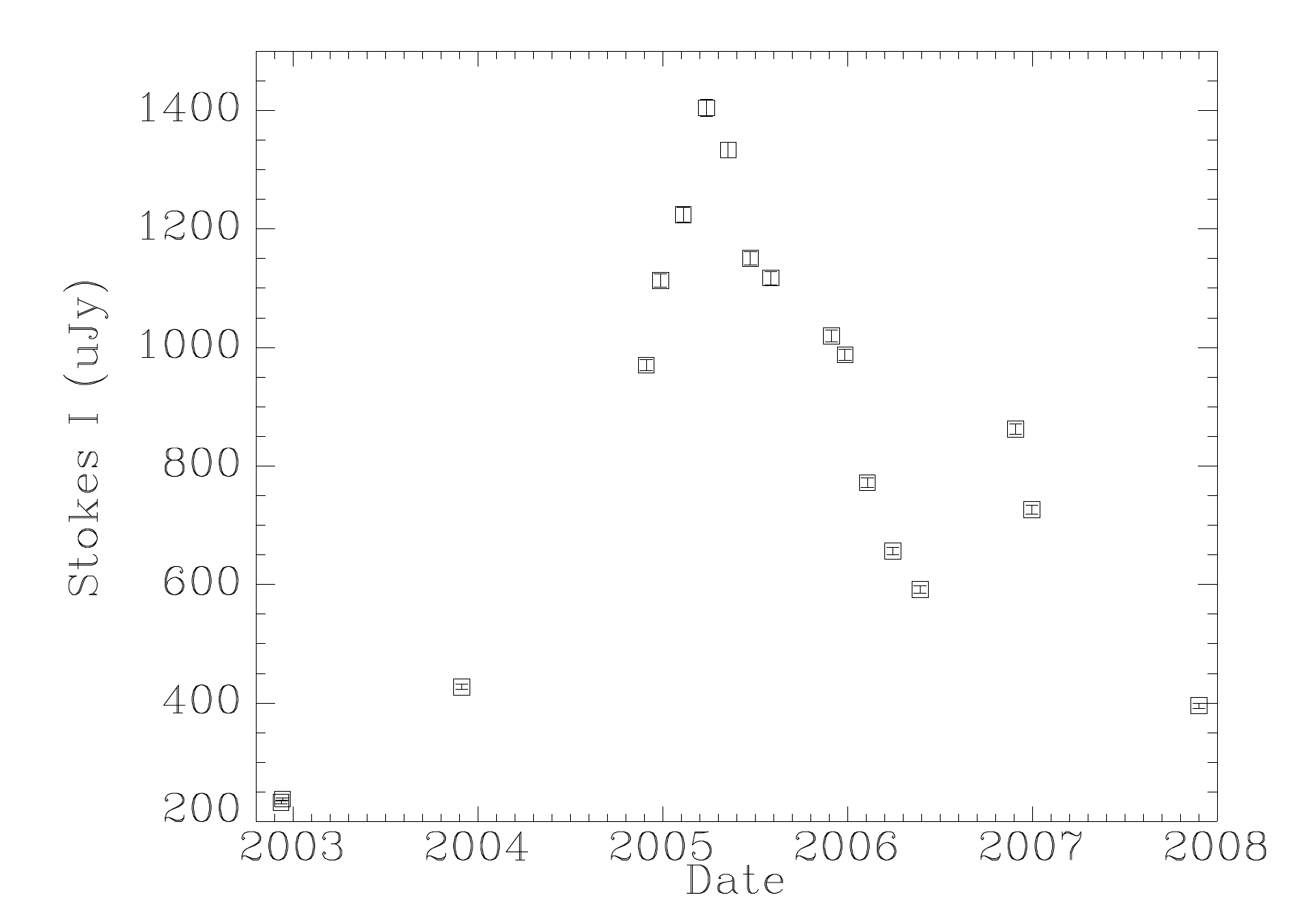}\\
  \includegraphics[width=6.3cm,height=4.5cm]{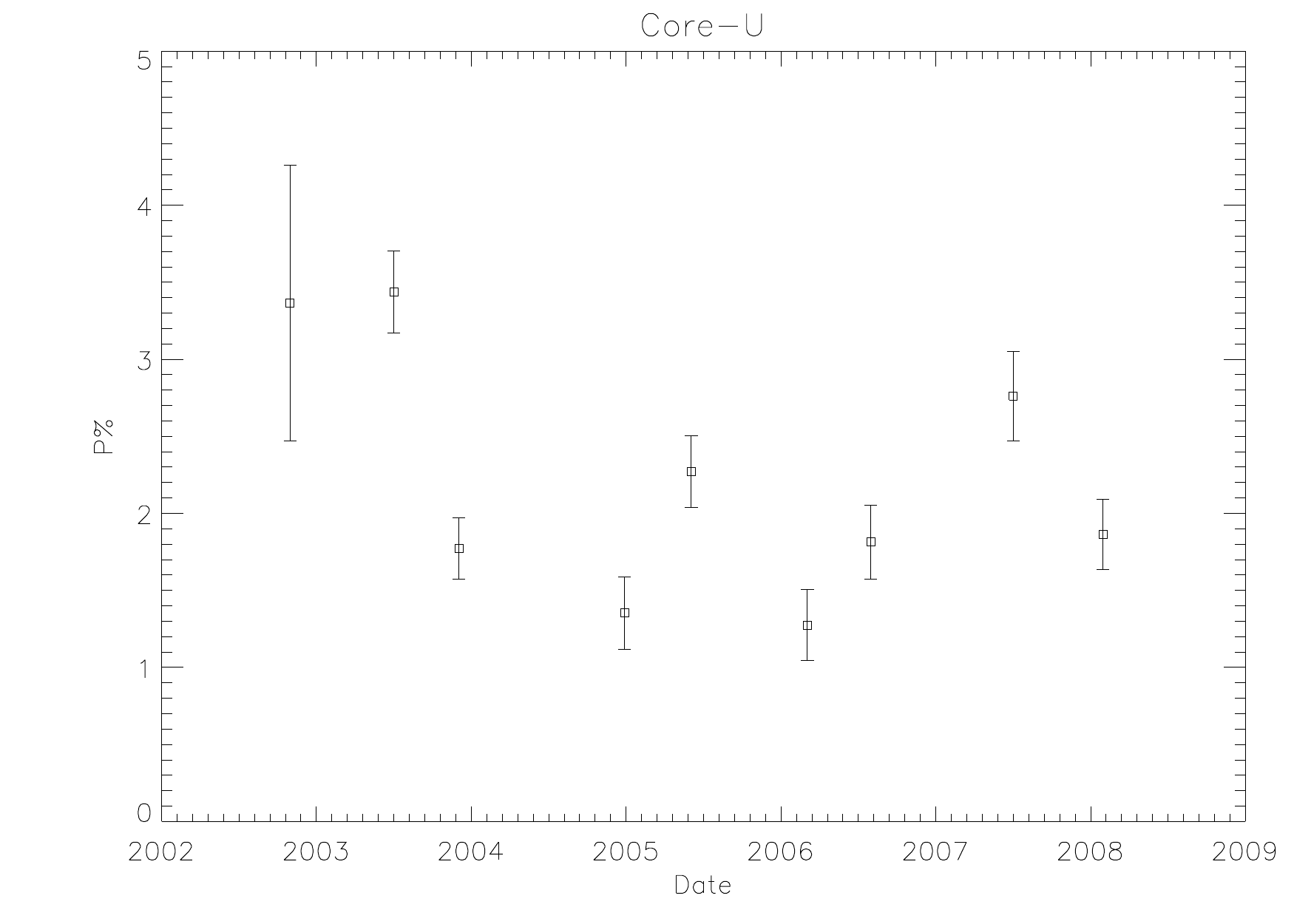}
  \includegraphics[width=6.9cm, height=4.5cm]{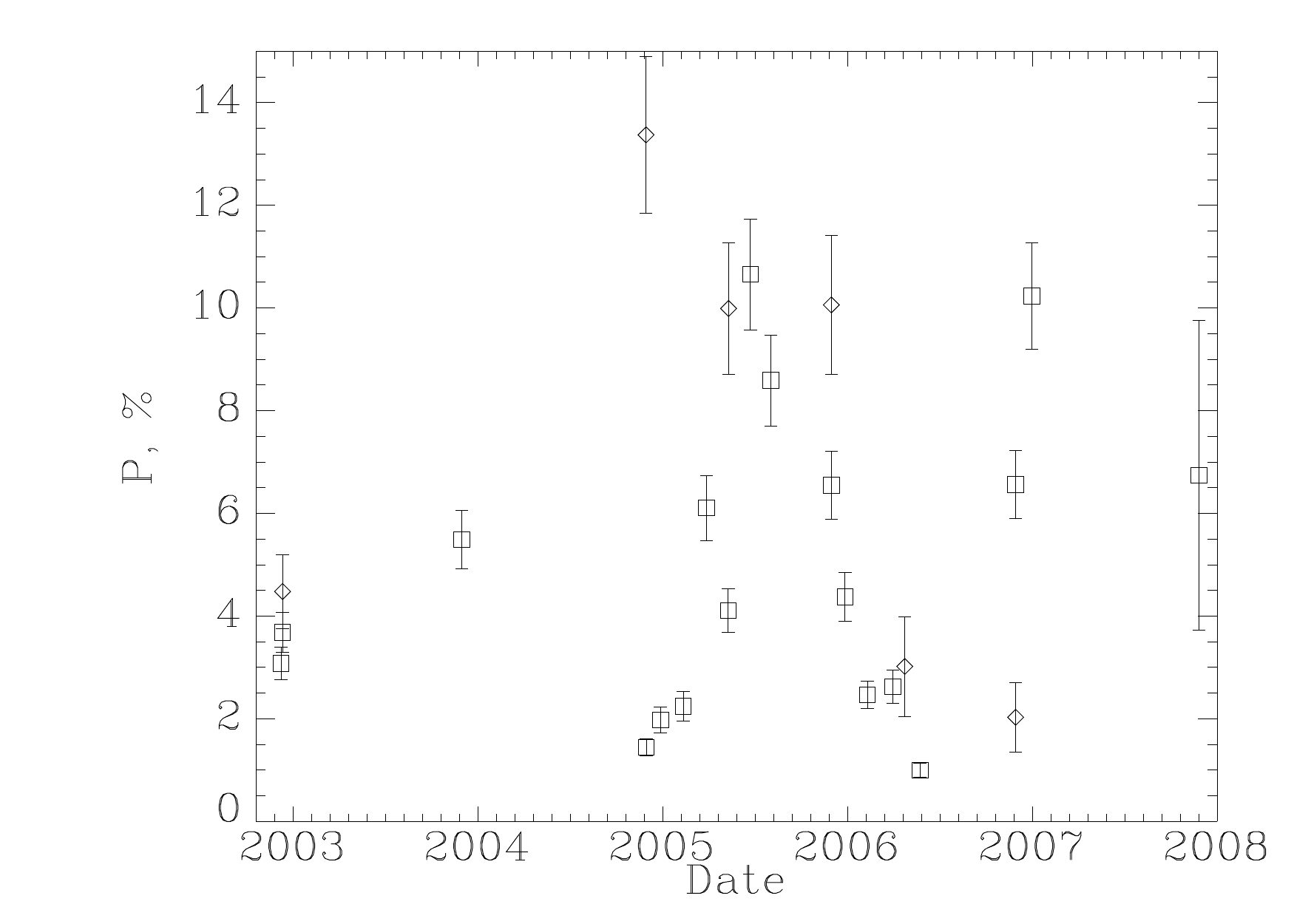}\\
  \includegraphics[width=6.2cm, height=4.5cm]{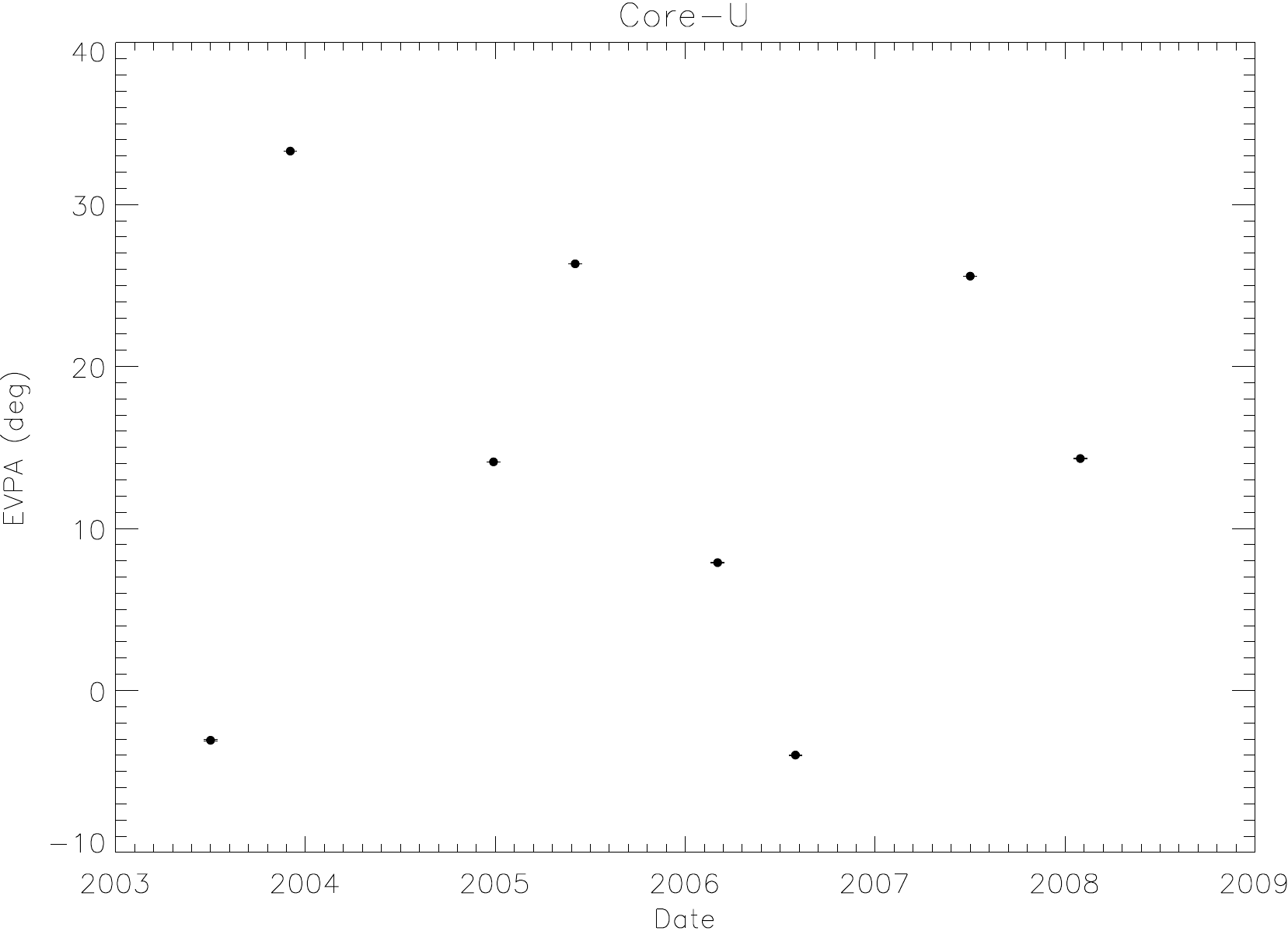}
  \includegraphics[width=6.9cm, height=4.5cm]{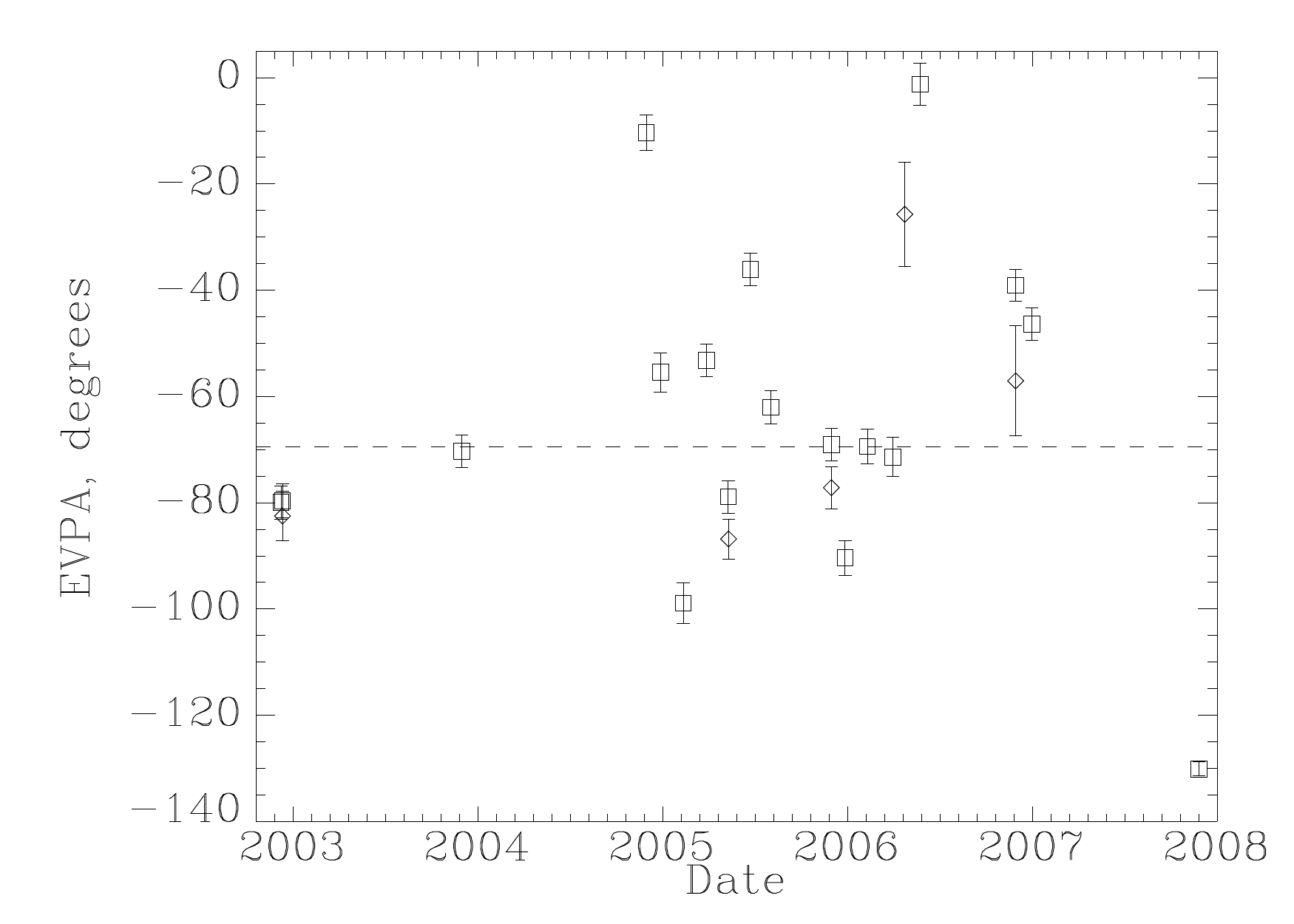}\\
   \end{tabular}
  \caption{Variability of total flux, percent polarization and EVPA. Left hand side panel shows radio (15 GHz) data: Top - the variability of the total flux of HST-1; middle - the variability of percent polarization of the nucleus and, bottom - the variability of EVPA of the nucleus with time. Right hand side panel shows the corresponding optical (F330W) plots (\cite[Perlman et al. 2011]{Perlman_etal11}).}
  \end{center}
  \label{fig3}
\end{figure}

\end{document}